\documentstyle[preprint,aps,]{revtex}

\begin{document}
 
\draft

\preprint{\begin{tabular}{r} KAIST-TH-98/02 \\ hep-ph/9808122 \end{tabular}}

\title{
Moduli Stabilization in Heterotic $M$-theory
}

\author{Kiwoon Choi\thanks{E-mail: kchoi@chep6.kaist.ac.kr},
Hang Bae Kim\thanks{E-mail: hbkim@supy.kaist.ac.kr}, and
Hyungdo Kim\thanks{E-mail: hdkim@supy.kaist.ac.kr}
}

\address{
Department of Physics, Korea Advanced Institute of Science and Technology \\
Taejon 305-701, Korea \\
}

\maketitle

\begin{abstract}
We examine the  stabilization of the two typical moduli, the 
length $\rho$ of the eleventh segment  and the volume $V$ of the
internal six manifold, in compactified heterotic $M$-theory. 
It is shown that, under certain conditions, the phenomenologically 
favored vacuum expectation values of $\rho$ and $V$ can be obtained 
by the combined effects of multi-gaugino condensations on the hidden 
wall and the membrane instantons wrapping the three cycle of
the internal six manifold.
\end{abstract}



Recently Ho\v{r}ava and Witten proposed that
the strong coupling limit of  the $E_8\times E_8$ heterotic
string theory can be described by the $d=11$ 
supergravity (SUGRA) on a manifold with boundaries
\cite{horava}. At energy scales below the $d=11$ 
Planck scale $M_{11}=\kappa^{-2/9}$, the 11-dimensional 
bulk and 10-dimensional boundary actions allow
a dimensional expansion in powers of $\kappa^{2/3}$
which may be interpreted as 
the inverse of the membrane tension \cite{foot1}.
Phenomenological implications of this heterotic
$M$-theory have been studied 
by compactifying the 11-dimensional SUGRA 
on a Calabi-Yau (CY) manifold times
the eleventh segment \cite{witten}.
It has been noticed that 
the resulting effective theory can  reconcile 
the observed $d=4$ Planck scale $M_{P}\approx
2.4 \times 10^{18}$ GeV 
with the phenomenologically favored unification
scale $M_{\rm GUT}\approx 3\times 10^{16}$
GeV in a natural manner \cite{witten,banks}, 
which was {\it not} possible in 
perturbative heterotic string theory.
In addition to providing a natural framework
for the unification of couplings,
heterotic $M$-theory  has other phenomenological virtues.
For instance, gaugino masses in heterotic $M$-theory 
appear to be comparable to the squark masses
even when  the $d=4$ supersymmetry is 
broken by  hidden sector gaugino condensation \cite{nilles}.
This is in contrast to the case of
perturbative heterotic string theory 
in which hidden sector gaugino condensation leads to 
undesirably small gaugino masses compared to 
the squark masses.
Another phenomenological virtue of $M$-theory 
is that there can be a QCD axion whose high energy axion potential is
suppressed enough so that the  strong CP problem can be solved 
by the axion mechanism \cite{banks,choi1}.

Compactified heterotic $M$-theory involves the two geometric moduli,
the length $\pi\rho$ of  the eleventh segment $S^1/Z_2$ and the 
volume $V$ of the internal six manifold  $X$ {\it averaged} over 
$S^1/Z_2$.
The above-mentioned  phenomenological virtues of  heterotic $M$-theory
have been discussed  based on the assumption that $\pi\rho$
and $V$ are stabilized
at the VEVs leading to the correct values of 
the $d=4$ gauge and gravitational couplings 
together with 
$M_{\rm GUT}\approx 3\times 10^{16}$ GeV. In this paper, 
we wish to study the stabilization of $\rho$ and $V$
in the context of $d=4$  effective action
including various nonperturbative effects,
e.g.  gaugino condensations on the hidden wall, 
membrane and fivebrane instantons \cite{Becker-Becker-Strominger}.
Our analysis indicates that
$\rho$ and $V$ can be stabilized at
the phenomenologically favored VEVs
by the combined effects of
multi-gaugino condensations and the membrane instantons
wrapping the three cycle ${\cal C}_3$ of $X$ if 
the hidden sector involves
multi-gauge groups with appropriate hidden matter contents,
and  $X$ admits
a complex structure for which the value of
$(32\pi^2)^{1/2}|\int_{{\cal C}_3}\Omega|/
(i\int_X \Omega\wedge\bar\Omega)^{1/2}$
is of order unity where $\Omega$ is the harmonic $(3,0)$ form on $X$.

To proceed, let us estimate the phenomenologically favored
values of $\langle\rho\rangle$ and $\langle V\rangle$.
To be definite, we will use the compactification involving a smooth
CY manifold.
Including the corrections at ${\cal O}(\kappa^{2/3})$,
the CY volume
is given by \cite{witten} 
\begin{equation}
V_{CY}(x^{11}) = V-(4\pi\kappa^2)^{1/3}\left(x^{11}-\frac12\right)
\pi\rho\int\omega\wedge I_4,
\label{moduli1}
\end{equation}
where $x^{11}$ covers $S^1/Z_2=[0,1]$
whose boundaries at $x^{11}=0$
and $x^{11}=1$ represent the visible
sector wall and the hidden sector wall, respectively,
$\pi\rho=\int dx^{11}\sqrt{g_{11 , 11}}$
is the physical length of $S^1/Z_2$,
$\omega$ is the CY K\"{a}hler form, and $I_4=\frac{1}{8\pi^2}[{\rm tr}
(F\wedge F-\frac{1}{2}R\wedge R)]$.  
Here we use the {\it downstairs} $d=11$ SUGRA coupling,
i.e. $\kappa^2\equiv
\kappa_{\rm down}^2=\frac{1}{2}\kappa_{\rm up}^2$ \cite{horava},
and also take into account the factor $2^{1/3}$ correction to the 
$d=10$ YM coupling made in 
\cite{harmark}.
Obviously, $V$ corresponds to the CY volume averaged over $S^1/Z_2$.
A simple dimensional reduction of the
$d=11$ bulk and $d=10$ boundary actions
leads to
the $d=4$  gauge coupling constant 
$\alpha_{\rm GUT}=(4\pi)^{2/3}\kappa^{4/3}\langle
V_{CY}(0)\rangle^{-1}$
at the unification scale
$M_{\rm GUT}=\langle V_{CY}(0)\rangle^{-1/6}$ 
and the $d=4$ Planck scale
$M_P=\kappa^{-1}\sqrt{\pi\rho V}$.
Fitting the phenomenological values of $\alpha_{\rm GUT}$, $M_P$ and
$M_{\rm GUT}$, one finds \cite{banks}
\begin{eqnarray}
\langle\pi\rho\rangle &\approx& 
15\,\kappa^{2/9} \approx (4\times10^{15}\,{\rm GeV})^{-1},
\nonumber\\
\langle V\rangle &\approx&
80\,\kappa^{4/3} \approx (3\times10^{16}\,{\rm GeV})^{-6}.
\label{vevs1}
\end{eqnarray}
For the gravitino mass $m_{3/2}={\cal O}(1)$ TeV,
the Kaluza-Klein scales  of compactified dimensions
are much higher than
the moduli masses which are presumed to be ${\cal O}(m_{3/2})$
and also than the dynamical scale of supersymmetry breaking,
e.g. the hidden gaugino condensation scale which would be
${\cal O}(10^{13})$ GeV.
This justifies our approach studying the moduli stabilization
in the context of the
$d=4$ effective SUGRA action.

In $d=4$ effective SUGRA,
$\rho$ and $V$ form
the chiral superfields $S$ and $T$
together with the axions arising from the 3-form gauge field in $d=11$ SUGRA.
Since $\langle \pi \rho\rangle$ is larger than 
$\langle V\rangle^{1/6}$ by one order of magnitude,
one can consider an intermediate $d=5$ effective theory
at length scales between $\langle V\rangle^{1/6}$
and 
$\langle \pi \rho\rangle$ \cite{banks,nilles,lukas}. 
In this scheme, at the leading order 
in $\kappa^{2/3}$,
$S$ belongs the $d=5$ hypermultiplet,
while $T$ and $d=4$ SUGRA multiplet belong to
the $d=5$ SUGRA multiplet. 
At any rate,  
$S$ and $T$ can be normalized 
by fixing the periodicity of their axion components:
\begin{equation}
{\rm Im}(S)\equiv {\rm Im}(S)+1, \quad
{\rm Im}(T)\equiv {\rm Im}(T)+1.
\label{normalization}
\end{equation}
In this normalization, 
\begin{eqnarray}
{\rm Re}(S) &=& (4\pi)^{-2/3}\kappa^{-4/3} V, \nonumber\\
{\rm Re}(T) &=& (4\pi\sum_{IJK}C_{IJK}/6)^{-1/3}\kappa^{-2/3}\pi\rho V^{1/3},
\end{eqnarray}
where $C_{IJK}=\int\omega_I\wedge\omega_J\wedge
\omega_K$ are the intersection numbers for 
$\omega_I$ $(I=1,..,h_{1,1})$ which form the basis 
of the integer $(1,1)$ cohomology.
To derive the second relation, we have used
$\kappa^{-2/3}
\pi\rho \, \omega 
=(4\pi)^{1/3} {\rm Re}(T)\sum_I\omega_{I}$
and  $V=\frac{1}{3!}\int
\omega\wedge\omega\wedge\omega$ for
the CY K\"ahler form $\omega$.
Then the moduli VEVs of Eq.~(\ref{vevs1}) correspond to  
\begin{eqnarray}
&&\langle {\rm Re}(S)\rangle \equiv
\langle S_R\rangle ={\cal O}(\alpha_{\rm GUT}^{-1}),
\nonumber\\
&&\langle {\rm Re}(T)\rangle \equiv
\langle T_R\rangle  ={\cal O}(\alpha_{\rm GUT}^{-1}),
\label{vevs2}
\end{eqnarray}
and our problem becomes to stabilize both $S_R$ and $T_R$ at the
VEVs of ${\cal O}(\alpha_{\rm GUT}^{-1})$.
To make a comparison, let us note that the dilaton and the overall
K\"ahler modulus in perturbative heterotic string theory 
are stabilized at
$\langle S_R\rangle={\cal O}(\alpha_{\rm GUT}^{-1})$
and $\langle T_R\rangle={\cal O}(1)$
when $S$ and $T$ are normalized as (\ref{normalization})
\cite{race}.

The moduli effective potential would be determined
by the K\"ahler potential ($K$)
and the superpotential ($W$) via the standard SUGRA formula: 
$V_{\rm eff}=e^{K}[K^{i\bar{j}}(D_iW)(D_jW)^*-3|W|^2]$.
Since we are interested in the possibility
for $\langle S_R\rangle \approx \langle T_R\rangle
={\cal O}(\alpha^{-1}_{\rm GUT})$,
let us consider the behavior of $K$ and $W$
in the limit $S_R\gg 1$ and $T_R\gg 1$.
In this limit, $K$ can be divided into two pieces,
$K=K_{\rm p}+K_{\rm np}$, where $K_{\rm p}$ is the part
which allows an asymptotic expansion in powers
of $1/S_R$ and $1/T_R$
and $K_{\rm np}$ stands for the rest
which originates from nonperturbative effects.
As was noted in \cite{Choi-Kim-Munoz}, $K_{\rm p}$
can be determined either in the context of the $M$-theory large
radius expansion which is available in the $M$-theory limit with
$\kappa^{-2/3}(\pi \rho)^3\approx \pi^2T_R^3/S_R
\gg 1$
or in the context of the string loop and $\sigma$-model
expansion which is available in the perturbative string limit 
$e^{2\phi}/(2\pi)^5
\approx \pi^2T_R^3/S_R \ll 1$.
(Here $\phi$ denotes  the heterotic string dilaton.)
Note that the asymptotic expansion of $K_{\rm p}$ in powers
of $1/S_R$ and $1/T_R$
is valid in both the $M$-theory limit and the perturbative string
limit as long as both $S_R$ and $T_R$
are large enough \cite{foot3}.
One key observation made in \cite{Choi-Kim-Munoz} is that, when
expanded in powers of 
$1/\pi S_R$ and $1/\pi T_R$,
the nonvanishing expansion coefficients are generically of order unity.
We then have
\begin{equation}
K = - \ln(S+\bar{S})-3\ln(T+\bar{T}) + \delta K_{\rm p} + K_{\rm np},
\label{Kahler-potential}
\end{equation}
where the leading logarithmic terms are determined
at the leading order in the $M$-theory large radius expansion
\cite{msugra} or
in the string loop and $\sigma$-model expansion, and 
the perturbative  corrections are  given by
$\delta K_{\rm p}=\sum_{n,m}C_{(n,m)}/(\pi S_R)^n(\pi T_R)^m$
with the coefficients $C_{(n,m)}$ which 
are essentially of order unity.
For the compactification on a CY space
with the minimal embedding, one finds 
$C_{(0,1)}=C_{(0,2)}=0$, $C_{(0,3)}=3\zeta(3)\chi/16\sum_{IJK}C_{IJK}$,
and $C_{(1,0)}=\chi/288$ \cite{Choi-Kim-Munoz},
showing that $C_{(n,m)}$ are indeed of order unity
(or less) for  reasonable values of the Euler number $\chi$ and the
intersection numbers $C_{IJK}$.
In fact, the most important corrections
with the coefficients $C_{(1,0)}$ and
$C_{(0,1)}$ can always be absorbed into the leading
logarithmic terms.
After absorbing these corrections, we have
\begin{equation}
\delta K_{\rm p}=
{\cal O}(1/(\pi S_R)^n(\pi T_R)^m)
={\cal O}((\alpha_{\rm GUT}/\pi)^2),
\end{equation}
for $n+m\geq 2$ and the moduli VEVs of (\ref{vevs2}). 
It is hard to imagine that such a small $\delta K_{\rm p}$
can play a significant role for the moduli stabilization, 
and so we will ignore it in the subsequent discussions.

As $M$-theoretic nonperturbative effects which may contribute to
$K_{\rm np}$, one can consider the following
types of instantons: $I_1=$ membrane instantons
wrapping the CY 3-cycle (${\cal C}_3$),
$I_2=$ membrane instantons which wrap the CY 2-cycle
(${\cal C}_2$) 
and are stretched along the 11-th segment,
$I_3=$ fivebrane instantons wrapping the entire CY volume.
These instantons 
have been discussed in \cite{Becker-Becker-Strominger} 
in the context of type IIA $M$-theory, however
it is rather straightforward to extend 
the discussion to the
heterotic $M$-theory. 
A complete computation of  the effects of these instantons
would require the full nonperturbative formulation of $M$-theory,
which is not available at this moment.
However one can still compute the most important semiclassical
factor $e^{-A(I)}$ where $A(I)$ is the Euclidean action of the instanton
$I$. 
A simple computation using
the membrane tension $T_2=(2\pi^2)^{1/3}\kappa^{-2/3}$ and
the fivebrane tension $T_5=(\pi/2)^{1/3}\kappa^{-4/3}$ yields
\cite{Becker-Becker-Strominger}
\begin{equation}
A(I_1)= b \sqrt{S_R}, \quad
A(I_2)= 2\pi k T,  \quad A(I_3)=2\pi S,
\label{instanton-action}
\end{equation}
where 
$k= |\sum_I\int_{{\cal C}_2}\omega_I|$ is a positive integer and
\begin{equation} 
b = (32\pi^2)^{1/2}\frac{|\int_{{\cal C}_3}
\Omega |}{(i\int \Omega\wedge \bar{\Omega})^{1/2}}
\end{equation}
for the harmonic $(3,0)$ form $\Omega$ on CY.
Note that generically $b$ is a function
of the complex structure moduli.
Since $A(I_2)$ and $A(I_3)$ are holomorphic,
$I_2$ and $I_3$ can affect not only the K\"ahler potential,
but also the holomorphic gauge kinetic functions and superpotential.
However $A(I_1)$ depends only on $S_R$, 
and thus the effects of $I_1$ can be encoded only through
the K\"ahler potential.

In fact, since their locations  
on $S^1/Z_2$ are not specified,  
$I_1$ and $I_3$
induce 5-dimensional local interactions
which are suppressed by
$e^{-{\cal A}(I)}$ where
${\cal A}(I_1)=b\sqrt{{\rm Re}({\cal S})}$
and ${\cal A}(I_3)=2\pi {\cal S}$
for the $d=5$ field
${\cal S}$ in the universal hypermultiplet
which is normalized as 
${\rm Re}({\cal S})=(4\pi)^{-2/3}\kappa^{-4/3} V_{CY}$.
For the CY volume $V_{CY}$ depending upon $x^{11}$ as Eq.~(\ref{moduli1}),
we have ${\rm Re}({\cal S})=S_R-n(x^{11}-\frac{1}{2})T_R$
where the integer $n=\sum_I\int\omega_I\wedge I_4$. 
After the reduction to $d=4$,
the 5-dimensional  interactions induced by $I_1$ and $I_3$ 
are {\it averaged} over $x^{11}$, yielding the 4-dimensional
local interactions suppressed by $e^{-A(I_{1,3})}$
with  $A(I_{1,3})$ given in (\ref{instanton-action}).

It is rather obvious that $I_2$ and $I_3$ are {\it irrelevant}
for the moduli stabilization at $\langle S_R\rangle
\approx \langle T_R\rangle=
{\cal O}(\alpha^{-1}_{\rm GUT})$ since their effects are suppressed
by the extremely small $e^{-A(I)}={\cal O}(e^{-2\pi/\alpha_{\rm GUT}})$.
As was noticed in \cite{Becker-Becker-Strominger}, 
$I_1$ can  generate a four-dilatino
operator, thereby modifying the Riemann-K\"ahler curvature tensor.
This would result in the nonperturbative correction 
$K_{\rm np}\propto e^{-A(I_1)}$.
A key feature distinguishing $I_1$ from $I_2$ and $I_3$  is that
it is {\it possible} that
$A(I_1)={\cal O}(1)$ even for $S_R={\cal O}(\alpha^{-1}_{\rm GUT})$,
which would be the case 
if $b={\cal O}(1)$
for some values of the complex structure moduli.
In this case, $I_1$  can give a sizable contribution
to $K_{\rm np}$ and thus be relevant for the moduli
stabilization.
However we stress that  
a rather particular  form of the complex
structure is required to have $b={\cal O}(1)$.
For instance,
$b=3^{-3/4}(32\pi^2)^{1/2}\approx 7.8$ for the simple $Z_3$ orbifold, 
yielding $A(I_1)\approx 35$ for $S_R\approx 20$.
In this case, the effects of $I_1$ 
would be too small to be
relevant for the moduli stabilization.
At any rate, the discussions
in the previous paragraph suggest
the following form of the nonperturbative 
K\"ahler potential:
\begin{equation}
K_{\rm np} = d \left(\frac{S_R}{4\pi}\right)^{p/2} e^{-b\sqrt{S_R}}
\left[1+{\cal O}\left(\frac{1}{\pi S_R},\frac{1}{\pi T_R}\right)\right]
+ {\cal O}(e^{-2\pi S},e^{-2\pi T}),
\label{nonperturbative-Kahler-potential}
\end{equation}
where $d$ and the integer $p$ are introduced to parameterize
the unknown parts of the $I_1$-induced K\"ahler potential.
We note that this membrane instanton-induced K\"ahler potential
corresponds to the $M$-theory realization of the
stringy nonperturbative effects which has been discussed 
by Shenker \cite{shenker} and later applied to the dilaton stabilization
in perturbative heterotic string vacua \cite{banks2}.

Let us now turn to the effective superpotential $(W)$
of $S$ and $T$.
Since $I_1$ does not affect $W$ and also
the effects of
$I_2$ and $I_3$ are suppressed by  
$e^{-2\pi T}$ and
$e^{-2\pi S}$ respectively, in the limit
$S_R\gg 1$ and $T_R\gg 1$,  $W$ is expected to be dominated by
the field-theoretic  
gaugino condensations.
More explicitly \cite{race},
\begin{equation}
W = \sum_a C_a e^{-\lambda_a f_a}
+{\cal O}\left(e^{-2\pi S},e^{-2\pi T}\right),
\label{superpotential}
\end{equation}
where $f_a$ denotes the gauge kinetic function of the
$a$-th hidden gauge group $G_a$, $\lambda_a$ is determined
by the one-loop beta function of $G_a$, and 
$C_a$ is also determined by 
$G_a$ and the hidden matter contents.
For instance, if $G_a=SU(N)$ and there are $M$ quarks
in $(N+\bar{N})$ representation, we have
$C_a= -(N-M/3)(32\pi^2e)^{\frac{M-N}{N-M/3}}
(M/3)^{\frac{M/3}{N-M/3}}$ and
$\lambda_a=8\pi^2/(N-M/3)$.

The gauge kinetic functions of the compactified 
heterotic $M$-theory are given by \cite{choi1,stieberger} 
\begin{equation}
4\pi f_a =
S -\frac{n_a}{2} T + {\cal O}(e^{-2\pi S},e^{-2\pi T})
\label{gauge-kinetic-function}
\end{equation}
where $n_a$ are model-dependent quantized coefficients.
For compactifications on a smooth CY times $S^1/Z_2$, 
$n_a=\frac{1}{8\pi^2}\sum_I\int \omega_I\wedge 
[{\rm tr}(F^2-\frac{1}{2}R^2)]={\rm integer}$,
however for compactifications involving a singular 
six manifold, e.g. orbifold, $n_a$ are generically rational
numbers depending upon the orbifold twists and also
the instanton numbers of the gauge bundle \cite{stieberger}.

{}From the K\"ahler potential
(\ref{Kahler-potential}) and (\ref{nonperturbative-Kahler-potential}),
the moduli effective potential is calculated to be
\begin{equation}
V_{\rm eff}=
\frac{e^{K_{\rm np}}}{16S_RT_R^3}
\left\{ \frac{\left|2S_RW_S-(1-\Delta)W\right|^2}{1+\Delta'}
+ \frac13\left|2T_RW_T-3W\right|^2 - 3|W|^2 \right\},
\label{potential}
\end{equation}
where $W_S=\partial W/\partial S$, $W_T=\partial W/\partial T$ and
$\Delta=\frac12(p-bS_R^{1/2})K_{\rm np}$,
$\Delta'=\frac14\{p(p-2)-(2p-1)bS_R^{1/2}+b^2S_R\}K_{\rm np}$.
When there is an appropriate minimum of the potential, we can calculate
the gravitino mass and also  the auxiliary
$F$-components of moduli which are given by
$m_{3/2}^2 = e^{K^{\rm np}}|W|^2/16S_RT_R^3$,
$F^S=2m_{3/2}S_R[2S_RW_S-(1-\Delta)W]/(1+\Delta')W$, and
$F^T=2m_{3/2}T_R[2T_RW_T-3W]/3W$.
We wish to examine whether
the potential (\ref{potential}) given by the superpotential
(\ref{superpotential})
can achieve a (local) minimum with the moduli VEVs of (\ref{vevs2})
and supersymmetry breaking with the weak scale gravitino mass,
for a reasonable choice of the hidden sector gauge group and matter fields,
and also of
the values of $d$, $p$, and $b$ describing the
membrane instanton-induced 
K\"ahler potential.
We will require that the moduli K\"ahler metrics are positive-definite 
over a sizable domain around $S_R\approx T_R={\cal O}(\alpha_{\rm GUT}^{-1})$,
however {\it not} require the moduli potential to  vanish at the minimum
since it does not correspond to the fully renormalized
vacuum energy density.
In this paper, we present some of the results of our analysis
to show the existence of the desired $M$-theory minimum,
and the full details will be presented elsewhere \cite{Choi-Kim-Kim}.

Let us first consider the case with  single gaugino condensation
yielding  $W=C_1e^{-\lambda_1 f_1}$ for the hidden sector gauge kinetic
function $4\pi f_1=S-\frac{n}{2}T$.
The value of $\langle S_R\rangle$
can be set to ${\cal O}(\alpha_{\rm GUT}^{-1})$ by a reasonable choice of
hidden sector gauge group and also
of the membrane-instanton-induced K\"ahler potential $K_{\rm np}$.
However, we have $\langle T_R\rangle=12\pi/n\lambda_1$
which can {\it not} be ${\cal O}(\alpha_{\rm GUT}^{-1})$
for any reasonable values of $\lambda_1$ and $n$,
particularly for the values which give rise to
the weak scale gravitino mass.
In fact, the minimum found in this case has 
$\langle S_R\rangle={\cal O}(\alpha_{\rm GUT}^{-1})$
and $\langle T_R\rangle={\cal O}(1)$, and thus corresponds to
the perturbative heterotic string vacuum discussed in \cite{race,banks2},
{\it not} the $M$-theory vacuum  that we are looking for.
We therefore conclude that even in the presence of a sizable
$K_{\rm np}$, single gaugino condensation does {\it not} lead to
the phenomenologically favored $M$-theory vacua with
the moduli VEVs (\ref{vevs2}) and the weak scale gravitino mass.

When there are two gaugino condensates with 
$W=C_1e^{-\lambda_1 f_1}+C_2
e^{-\lambda_2 f_2}$ and also a sizable 
$K_{\rm np}$, it turns out that
the potential can have  the desired $M$-theory minimum.
One may first consider a superpotential implementing 
the simplest form of the $T$-duality, $T\rightarrow
1/T$:
$W=\eta^{-6}(T)(C_1 e^{-\lambda_1 S/4\pi}+C_2 e^{-\lambda_2 S/4\pi})$.
However this type of superpotential always  leads to  
$\langle T_R\rangle={\cal O}(1)$ and thus
the perturbative heterotic string vacuum, {\it not} 
the $M$-theory vacuum.
Motivated by the results in CY cases,
here we  consider an alternative simple case that the two hidden
gauge groups have the same gauge kinetic function:
$4\pi f_1=4\pi f_2=S-\frac{1}{2}T$, while the visible
sector gauge kinetic function is $4\pi f_{\rm v}=S+\frac{1}{2}T$.
In this case, the two gaugino condensations fix the VEV of
${\rm Im}(S-\frac{1}{2}T)$ to be
$4\pi^2l/(\lambda_1-\lambda_2)$ where $l$ is an odd integer
and also can stabilize $S_R-\frac{1}{2}T_R$ by the conventional race-track
mechanism \cite{race}.
In the absence of a sizable $K_{\rm np}$,
the moduli potential still has a run-away behavior along
$S_R+\frac{1}{2}T_R$. 
However with a proper membrane instanton-induced $K_{\rm np}$,
a  minimum with the desired moduli VEVs~(\ref{vevs2}) can be formed.
This minimum is located  in the valley of the potential
which is formed because the two curves of $F^S=0$ and $F^T=0$ come close.
We could  actually find several examples which give rise to 
the desired moduli VEVs (\ref{vevs2})
for the reasonable values of parameters, which are shown in Table~1.
Note that the solution to $F^S=F^T=0$ is always an extremum point,
however in our example it turns out to be a saddle point.
Without the hidden matter fields, the minimum is located
at near $S_R-\frac{1}{2}T_R=0$,
which would result in a too large gravitino mass.
If appropriate hidden matter fields are assumed,
we can get the minimum with 
the desired moduli VEVs (\ref{vevs2}) and the weak scale gravitino mass.
For the examples depicted in Table 1,
supersymmetry breaking is characterized by $F^S\approx F^T$
which may lead to an interesting pattern of soft terms \cite{Choi-Kim-Kim}.

In conclusion, we have examined the stabilization of the
two typical $M$-theory moduli,
the length $\rho$ of the 11-th segment $S^1/Z_2$ and the volume
$V$ of the internal six manifold $X$ averaged over $S^1/Z_2$.
A particular attention was paid for the
possibility that these moduli are stabilized at the VEVs
which give rise to the correct values
of the 4-dimensional gauge and gravitational coupling constants
together with $M_{\rm GUT}\approx 3\times 10^{16}$ GeV.
Such moduli VEVs
could be obtained by the combined effects of
multi-gaugino condensations and the membrane instantons
wrapping the three cycle ${\cal C}_3$ of $X$ if 
the hidden sector involves
multi-gauge groups with appropriate hidden matter contents,
and  $X$ admits a complex structure for which the value of
$(32\pi^2)^{1/2}|\int_{{\cal C}_3}\Omega|/
(i\int_X \Omega\wedge\bar\Omega)^{1/2}$
is of order unity where $\Omega$ is the harmonic $(3,0)$ form on $X$.

\section*{Acknowledgments}
This work is supported in part by KOSEF, 
through CTP of Seoul National University,
KRF under the Distinguished Scholar Exchange Program,
and Basic Science Institute Program BSRI-97-2434.

\begin{table}
\newcommand{\ten}[2]{#1\times10^{#2}}
$$\begin{array}{|cccc|ccc||c|c|c|}
\hline
N_1&M_1&N_2&M_2&d&p&b&\langle S_R \rangle&\langle T_R \rangle&m_{3/2}({\rm
GeV})\\\hline\hline
3&0&4&8&8&8&0.5&19&16&\ten{3.9}{2} \\
3&0&4&8&2&12&1&19&17&\ten{3.9}{2} \\
3&0&4&8&8&16&1.5&19&16&\ten{4.3}{2} \\
3&1&4&10&2.3&12&1&19&17&\ten{9.6}{2} \\
3&2&4&11&2.5&12&1&18&17&\ten{7.1}{2} \\ \hline
\end{array}$$
\caption{Moduli VEVs for 
the hidden gauge group
$SU(N_1)\times SU(N_2)$ with the hidden matters
$M_1(N_1,1)+M_2(1,N_2) +{\rm c.c.}$ and 
$4\pi f_1=4\pi f_2=S-\frac{1}{2}T$.}
\end{table}


\begin{thebibliography}{99}
\bibitem{foot1}
Note that all other dimensionful parameters, e.g. the
bulk SUGRA coupling ($\kappa^2$),
the boundary YM coupling ($\kappa^{4/3}$), and also
the fivebrane tension $(\kappa^{-4/3}$), appear as some 
powers of $\kappa^{2/3}$. As a result, counterterms induced
by the loops involving the SUGRA and/or YM couplings and also
those induced by fivebranes can be properly taken into account
in the expansion in powers of $ \kappa^{2/3}$.
\bibitem{foot3}
Although not phenomenologically interesting,
$M$-theory moduli space includes  a  domain on
which $\kappa^{2/9}\ll \pi\rho\ll V^{1/6}$.
On this domain,
starting from the $d=11$ bulk and $d=10$ boundary actions
expanded in powers of $\kappa^{2/3}$, 
one can systematically
compute the $d=10$ action which is valid at
the energy scales below $(\pi\rho)^{-1}$ but still
above $V^{-1/6}$.
In this $d=10$ action,
the original $\kappa^{2/3}$-expansion
is splitted into the double expansion
in powers of $\kappa^{4/3}$ and $\kappa^{2/3}/\pi \rho$,
which corresponds to 
the string loop expansion in $e^{-2\phi}(\alpha^{\prime})^3
\sim \kappa^{4/3}$ and the $\sigma$-model
expansion in $\alpha^{\prime}\sim \kappa^{2/3}/\pi\rho$
when this action is extrapolated 
to the perturbative string vacua.
If one goes further down  to the energy scales  below $V^{-1/6}$,
the $d=4$ action appears to allow the double expansion
in powers of $1/S_R\sim \kappa^{4/3}/V$ and $1/T_R\sim \kappa^{2/3}/\pi\rho
V^{1/3}$.
Note that the $M$ and string theory duality implies that
the string loop expansion in the $d=10$ action
corresponds to a dimensional expansion
involving $(\alpha^{\prime})^3$.
This  leads to for instance  the nonrenormalization of $R^{(1+3k)}$
by the string $l(\geq k+1)$-loop effects where
$R$ denotes the $d=10$ Riemann curvature tensor.
\end{thebibliography}
\end{document}